\documentclass[aps,prl,twocolumn,superscriptaddress]{revtex4-1}
\usepackage{amsmath,amssymb,graphicx,natbib,aas_macros,bm}
\usepackage{enumerate}
\usepackage{bbold}
\usepackage{hyperref}
\usepackage{breakurl}

\usepackage{tabularx}
\usepackage{bigstrut}
\usepackage{multirow}

\newcommand{\GeV}{\ensuremath{\mathrm{GeV}}}

\begin{document} 

\title{Dark Matter from Minimal Flavor Violation}

\author{Brian Batell}
\email{bbatell@perimeterinstitute.ca}
\affiliation{Perimeter Institute for Theoretical Physics, Waterloo,
ON, N2L 2Y5, Canada}

\author{Josef Pradler}
\email{jpradler@perimeterinstitute.ca}
\affiliation{Perimeter Institute for Theoretical Physics, Waterloo,
ON, N2L 2Y5, Canada}

\author{Michael Spannowsky}
\email{mspannow@uoregon.edu}
\affiliation{Institute of Theoretical Science, University of Oregon, Eugene, OR 97403-5203, USA
}

\begin{abstract}
  We consider theories of flavored dark matter, in which the dark
  matter particle is part of a multiplet transforming nontrivially
  under the flavor group of the Standard Model in a manner consistent
  with the principle of Minimal Flavor Violation (MFV).  
  MFV automatically leads to the stability of the lightest state for a
  large number of flavor multiplets. If neutral, this particle is an
  excellent dark matter candidate. Furthermore, MFV implies specific
  patterns of mass splittings among the flavors of dark matter and
  governs the structure of the couplings between dark matter and
  ordinary particles, leading to a rich and predictive cosmology and
  phenomenology. We present an illustrative phenomenological study 
  of an effective theory of a flavor $SU(3)_Q$ triplet, gauge singlet scalar.
\end{abstract}

\maketitle

The observational evidence for dark matter (DM) provides one of the
few solid empirical motivations for new particle
physics~\cite{review}. Indeed, the simplest explanation of DM is a new 
massive particle that is stable on cosmological time scales and a
thermal relic of the Big Bang. Many scenarios for physics beyond the
Standard Model (SM) include candidates for the DM particle, with a
popular example being the neutralino in theories with weak scale
supersymmetry (SUSY).

Typically, in such theories the stability of the DM particle is a
consequence of a discrete $Z_2$ parity symmetry motivated by the need
to eliminate dangerous operators allowed by SM gauge invariance. For
example, in the Minimal Supersymmetric Standard Model (MSSM)
renormalizable terms in the superpotential can violate baryon and
lepton number and thus lead to rapid proton decay. R-parity
conservation is often invoked to eliminate such operators, with the welcome
side effect of rendering the lightest supersymmetric particle stable
and therefore a DM candidate.

It is remarkable that certain phenomenological constraints such
as proton decay may be overcome by invoking discrete symmetries, which
in turn lead to stable DM candidates. However, our detailed knowledge
of the flavor sector of the SM typically requires a highly 
non-trivial structure in the couplings of new particles to the quarks
and leptons---one that is not easily achieved by imposing discrete
symmetries. For example, R-parity does not prevent large flavor
changing neutral currents (FCNCs) in the MSSM, and additional
assumptions about the structure of SUSY breaking must be made.
In this regard the principle of Minimal Flavor Violation (MFV) is well
motivated~\cite{MFV1,MFV2}. One postulates that the global flavor
symmetry of the SM is broken only by the Yukawa matrices in the
presence of new particles. This provides a built-in protection
mechanism against large flavor violating processes. We shall
demonstrate that, in analogy with the discrete symmetries discussed
above, the MFV principle can lead to stable neutral particles and
excellent DM candidates.

In this paper we propose new theories of flavored DM. We add to the SM
new matter multiplets with nontrivial transformation properties under
the flavor group. We show that the MFV hypothesis implies that certain
flavor representations contain one or more stable components, and we
identify the general conditions for stability in these theories. Many
variants of flavored DM models can then be constructed, each with a
rich and predictive phenomenology.

As an example, we study a simple effective theory of DM with a flavor
$SU(3)_Q$ triplet, gauge singlet scalar field, which provides a
representative survey of various aspects of the cosmology and
phenomenology in the flavored DM scenario. MFV predicts patterns in
the mass splittings of the DM flavors governed by the SM Yukawa
couplings.  Flavor is intertwined with the physics of DM, as the
Yukawa couplings and CKM matrix play a role in the thermal freezeout,
direct and indirect signatures, and collider signals.  In the scenario
studied here, the heavy flavors in the DM multiplet decay to lighter
flavors along with SM particles, leading to striking signatures at
collider experiments. In other models, it is possible that the heavy
flavors are  stable on cosmological time scales, in which case
several species comprise the cosmic DM.

We note that several related studies exist in the literature.  As an alternative to R-parity,
Ref.~\cite{Nikolidakis:2007fc} considers MFV as a protection mechanism
in the MSSM, although in this scenario the LSP is not stable on
cosmological times scales and is therefore not a DM candidate.  Supersymmetric DM tied to the
neutrino flavor structure has been considered
in~\cite{MarchRussell:2009aq} while Ref.~\cite{kile} 
investigated theories of flavored DM, though both of these proposals are outside
the MFV framework.
Lastly, and in a more general context, models of flavor employing
\textit{e.g.} discrete symmetries~\cite{discrete1,discrete2} or horizontal
gauge symmetries~\cite{horizontal} have also found their application
in the DM problem.

\paragraph{\bf Flavor symmetries and dark matter.}

We begin by reviewing the MFV hypothesis, which provides a mechanism
based on flavor symmetries by which new light particles may be
consistent with precision flavor observables.  The quark sector of SM
exhibits a large global flavor symmetry $G_q = SU(3)_Q \times
SU(3)_{u_R} \times SU(3)_{d_R}$, which is broken only by Yukawa
interactions.  When considering theories of physics beyond the SM, the
MFV principle dictates that new particles also respect $G_q$ and the
only sources of flavor breaking arise from insertions of the SM Yukawa
matrices $Y_u$ and $Y_d$~\cite{MFV1,MFV2}.  Formally, this is achieved
by promoting the Yukawa matrices to spurion fields which transform
nontrivially under $G_q$. The SM Yukawa interactions are written as
\begin{eqnarray}
 -{\cal L}_Y & \supset &  \bar Q  Y_d  d_R H  + \bar Q Y_u u_R H^\dag + {\rm h.c.}  ,
\label{yukawa}
\end{eqnarray}
so that the Yukawa spurions transform as $Y_u \sim ( \bf{3} , {\bf
  \bar 3} , {\bf 1} )$, $Y_d \sim ( \bf{3} , {\bf 1} , \bf{ \bar 3} )$
under 
$G_q$.

A number of studies considering new MFV matter content exist in the
literature~\cite{Manohar:2006ga, Grossman:2007bd,Bauer:2009cc,
  Arnold:2009ay,Gross:2010ce,Arnold:2010vs}.  In all of these works,
the $G_q$ representations of the new particles have been chosen such
that renormalizable tree-level couplings to the SM fermions are
allowed. An immediate consequence of this is that such particles are
unstable and decay to SM quarks or leptons. We now investigate the
generality of this conclusion, and show in particular that MFV implies
stability for particles in certain representations of $G_q$. With
further assumptions about their electroweak quantum numbers, these
stable particles become perfectly viable DM candidates.

Let $\chi$ be a matter multiplet that transforms non-trivially under
flavor but is color-neutral, \textit{i.e.}~a singlet under
$SU(3)_c$. The most general operator that induces the decay of $\chi$
then reads
\begin{eqnarray}
  \label{decayOP}
  {\cal O}_{\rm decay} & = & 
  \chi \, 
  \underbrace{ Q \ldots }_{\mbox{ $A$ }}
  \underbrace{ \bar Q \ldots }_{\mbox{ $B$ }}
  \underbrace{ u_R \ldots }_{\mbox{ $C$ }}
  \underbrace{ \bar u_R \dots }_{\mbox{ $D$ }}
  \underbrace{ d_R \dots }_{\mbox{ $E$ }}
  \underbrace{ \bar d_R \dots }_{\mbox{ $F$ }} ~~~~~~~~~~ \\ 
  && ~~~~~~~~~~~~~\times
  \underbrace{ Y_u \dots }_{\mbox{ $G$ }}
  \underbrace{ Y^\dag_u \dots }_{\mbox{ $H$ }}
  \underbrace{ Y_d \dots }_{\mbox{ $I$ }}
  \underbrace{ Y^\dag_d \dots }_{\mbox{ $J$ }} \, {\cal O}_{\mathrm{weak}}, \nonumber
\end{eqnarray} 
where $A$ is the number of $Q$ fields, $B$ is the number of $\bar Q$
fields, etc.  $ {\cal O}_{\mathrm{weak}}$ is a potential electroweak
operator so that (\ref{decayOP}) is rendered invariant under
$SU(2)_L\times U(1)_Y$.
The flavor and color indices of the remaining fields are to be contracted by
use of the $SU(3)$ group invariant tensors $\delta$ and $\epsilon$ so
that (\ref{decayOP}) becomes a color and flavor singlet.
This, however, can only be achieved if the triality $t_i $ of each $SU(3)_i$
tensor product $(p,q)_i$ with $p$ factors of $\mathbf{3}_i$ and $q$
factors of $\mathbf{\bar3}_{i}$, vanishes for the operator ${\cal O}_{\rm decay}$:
\begin{equation}
  \label{eq:trialityCond}
  t_i  \equiv
  (p-q)_i \,{\rm mod}\,3 =0,  \quad i = c,\,Q,\,u_R,\, d_R.
\end{equation}

On the contrary, the decay operator (\ref{decayOP}) will be forbidden
if $t_i \neq 0$ for at least one $i$. Denoting the irreducible
representation of $\chi$ under $G_q$ as
\begin{equation}
\chi \sim (n_Q, m_Q)_{Q}\times (n_u, m_u)_{u_R}\times (n_d, m_d)_{d_R},
\end{equation}
where $n_Q$, $m_Q$, etc. can take values $0,1,2,\dots$, the
triality conditions (\ref{eq:trialityCond}) become
\begin{align}
  t_c & =( A \!-\! B\!  +\! C\! -\! D\!  +\! E\! - \!F )\:{\rm mod}\:3 =0  ,\label{t1}   \\
  t_Q & = (n_Q\! -\! m_Q\! +\!  A\! -\! B\! +\! G\! -\! H\! +\! I\! -\! J  )\:{\rm mod}\:3 =0 ,\label{t2}\\
  t_{u_R} & = (n_u \!- \!m_u \!+ \! C\! -\! D\! -\! G\! + \!H   )\:{\rm mod}\:3  =0  ,\label{t3} \\
  t_{d_R} & = (n_d \!- \!m_d \!+ \! E\! -\! F\! -\! I\! + \!J )\:{\rm
    mod}\:3 =0  .\label{t4}
\end{align}
Adding together Eqs.~(\ref{t2}-\ref{t4}) and subtracting
Eq.~(\ref{t1}), we find that a necessary condition for ${\cal O}_{\rm decay}$ to be allowed, and thus for $\chi$ to be unstable is  $(n-m) \:{\rm mod}\:3 = 0$, where $n \equiv n_Q + n_u + n_d$ and $m \equiv m_Q + m_u + m_d$. It follows that that ${\cal O}_{\rm decay}$ is forbidden and $\chi$ is stable if
\begin{align}
\label{stable}
(n-m) \:{\rm mod}\:3 \neq 0 .
\end{align}
Once (\ref{stable}) holds, $\chi$ contains a stable component.  In
this regard, it is important to note that for operators with multiple
fields $\chi$, $\chi^\dag$ which may potentially mediate a loop
induced decay, the stability condition above still holds. 
Table~\ref{DMcandidates} lists the lowest-dimensional
representations of $G_q$ that are stable according to the
condition~(\ref{stable}).

In order to provide a viable theory of DM, any stable state in the
flavored multiplet $\chi$ must further be electrically neutral. It
thus remains to specify the electroweak quantum numbers of $\chi$.
One possibility is that $\chi$ is a SM gauge singlet~\cite{singlet}. Alternatively,
as discussed in Ref.~\cite{minimalDM}, $\chi$ may be a 
$n$-plet of $SU(2)_L$ with hypercharge $Y$ such that a component of
$\chi$ is neutral, $Q = T_3 + Y = 0$, with $T_3$ being the diagonal
$SU(2)_L$ generator. According to this condition, a $SU(2)_L$ doublet
has hypercharge $Y=\pm 1/2$, a $SU(2)_L$ triplet has hypercharge
$Y=0,\pm 1$, and so on.

\begin{table}[t]
\begin{center}
\renewcommand{\arraystretch}{1.5}
 \begin{tabular}{| c || c | c |  }
\hline
 $(n,m)$ & ~ $SU(3)_Q\times SU(3)_{u_R}\times SU(3)_{d_R}$~ & ~Stable?~       \\
 \hline \hline
  $(0,0)$ & $ ( {\bf 1}, {\bf 1} ,{\bf 1} )$   &       \\ \hline
 $(1,0)$ & $ ( {\bf 3}, {\bf 1} ,{\bf 1} )$,  
$ ( {\bf 1}, {\bf 3} ,{\bf 1} )$,  
$ ( {\bf 1}, {\bf 1} ,{\bf 3} )$   &  Yes     \\ \hline
$(0,1)$ & $ ( {\bf \bar 3}, {\bf 1} ,{\bf 1} )$,  
$ ( {\bf 1}, {\bf \bar 3} ,{\bf 1} )$,  
$ ( {\bf 1}, {\bf 1} ,{\bf \bar 3} )$   &  Yes     \\ \hline
\multirow{2}{*}{(2,0)} 
&  $ ( {\bf 6}, {\bf 1} ,{\bf 1} )$,  $ ( {\bf 1}, {\bf 6} ,{\bf 1} )$,  $ ( {\bf 1}, {\bf 1} ,{\bf 6} )$ &
\multirow{2}{*}{Yes} 
 \\ 
&  $ ( {\bf 3}, {\bf 3} ,{\bf 1} )$, $ ( {\bf 3}, {\bf 1} ,{\bf 3} )$, $ ( {\bf 1}, {\bf 3} ,{\bf 3} )$ & 
\multirow{2}{*}{}    
\\  \hline
\multirow{2}{*}{(0,2)} 
&  $ ( {\bf \bar 6}, {\bf 1} ,{\bf 1} )$,  $ ( {\bf 1}, {\bf \bar 6} ,{\bf 1} )$,  $ ( {\bf 1}, {\bf 1} ,{\bf \bar 6} )$ &
\multirow{2}{*}{Yes} 
 \\ 
&  $ ( {\bf \bar 3}, {\bf \bar 3} ,{\bf 1} )$, $ ( {\bf \bar  3}, {\bf 1} ,{\bf \bar 3} )$, $ ( {\bf 1}, {\bf \bar 3} ,{\bf \bar 3} )$ & 
\multirow{2}{*}{}    
\\  \hline
\multirow{3}{*}{(1,1)} 
&  $ ( {\bf  8}, {\bf 1} ,{\bf 1} )$,  $ ( {\bf 1}, {\bf  8} ,{\bf 1} )$,  $ ( {\bf 1}, {\bf 1} ,{\bf  8} )$ &
\multirow{3}{*}{} 
 \\ 
&  $ ( {\bf  3}, {\bf \bar 3} ,{\bf 1} )$, $ ( {\bf  3}, {\bf 1} ,{\bf \bar 3} )$, $ ( {\bf 1}, {\bf  3} ,{\bf \bar 3} )$ & 
\multirow{3}{*}{}    
\\  
&  $ ( {\bf  \bar 3}, {\bf 3} ,{\bf 1} )$, $ ( {\bf \bar  3}, {\bf 1} ,{\bf 3} )$, $ ( {\bf 1}, {\bf  \bar 3} ,{\bf  3} )$ & 
\multirow{3}{*}{}    
\\  \hline
    \hline
  \end{tabular}
\end{center}
\caption{ Flavored DM candidates. Listed are the lowest-dimensional
  representations of $G_q$, organized according $(n,m)$ where $n \equiv n_Q+n_u+n_d$, 
  $m\equiv m_Q + m_u + m_d$. We have also indicated the representations that are stable 
  once MFV is imposed. Depending on their electroweak quantum numbers, these multiplets 
  may contain viable DM candidates.
}
\label{DMcandidates}
\end{table}

Note that the logic leading to the stability condition (\ref{stable})
made no assumption about renormalizability. Therefore, the DM candidates are absolutely stable if we strictly enforce MFV, even at the nonrenormalizable level. We also made no assumptions about the Lorentz quantum numbers of $\chi$, so one may consider scalars, fermions, vectors, and so on. 
Furthermore,
if additional particle content is added besides the DM candidate our
stability condition holds provided we demand that $\chi$ contains the
lightest mass eigenstate among the new particles.

There exist variations in the precise implementation of
MFV. For example, in the case of `linear MFV', one imagines that each
Yukawa insertion comes with a small coefficient, so that only the
leading terms are important, whereas in `general MFV' all Yukawa
insertions are considered to be of similar size~\cite{generalMFV}. The
number of Yukawa spurions in Eq.~(\ref{decayOP}) was arbitrary, so
that the stability condition (\ref{stable}) holds for general MFV. 
The case of linear MFV is more restrictive and may thus lead to
additional DM candidates.

Accounting for the different possible Lorentz, gauge, and flavor
quantum numbers, we conclude that the MFV principle leads to a wide
variety of DM candidates. 
MFV thus provides an attractive alternative to other well-known stabilization symmetries, such as the canonical $Z_2$ parity, as well as higher Abelian or non-Abelian discrete or continuous symmetries~\cite{other}. To be viable, these models must pass a
number of constraints from cosmology and experiment. We now discuss
these issues in an example model with a flavored DM candidate.

\paragraph{\bf Flavor \boldmath$SU(3)_Q$ triplet dark matter.}
To illustrate in more detail the general phenomenological
considerations in the flavored DM scenario, here we examine in some
detail the physics of a simple flavored DM candidate: a $SU(3)_Q$
triplet, SM gauge singlet scalar field. We consider an effective
theory with higher dimensional operators coupling the DM to quarks,
and examine the constraints coming from cosmology, direct and indirect
detection probes, and cosmology. We shall see that the theory has a
rich phenomenology and leads to novel phenomena at hadron colliders.

For concreteness, we consider a model with a scalar field with the
following $SU(3)_c\times SU(2)_L\times U(1)_{Y}\times G_q$  quantum numbers:
\begin{equation}
S \sim ({ \bf 1},{\bf 1 }, {0 })_{\rm SM} \times ({ \bf 3},{\bf 1 }, {\bf 1 })_{G_q}  .
\end{equation}
The Lagrangian is
\begin{equation}
\mathcal{L} = \partial^\mu S^*_i \partial_\mu S_i - V(S_i, H)  + {\cal L}_{\rm eff},
\end{equation}
where $i=1,2,3$ is a flavor 
index, $V(S_i, H)$ is the scalar
potential involving $S_i$ and the SM Higgs $H$, and $\mathcal{L}_{\rm
  eff}$ contains higher dimensional operators involving two $S$ fields
and two quark fields.

\paragraph{Spectrum.} Let us first consider the scalar potential,
which determines the mass spectrum of the flavored DM multiplet. The
scalar $S$ may have a bare mass term as well as contribution from
electroweak symmetry breaking due to a Higgs portal coupling~\cite{singlet}. The
relevant terms in the potential are
\begin{eqnarray}
V& \supset & m_S^2 S^*_i (a\,\mathbf{1}_{ij}+ b\, (Y_u Y_u^\dag)_{ij}+\dots) S_j 
\\&+& 
2 \lambda \, S^*_i (a' \mathbf{1}_{ij}+ b' (Y_u Y_u^\dag)_{ij}+\dots)S_j\, H^\dag H  \nonumber, 
\label{portal}
\end{eqnarray}
where $a,b,a',b'$ are dimensionless MFV parameters. The ellipsis
indicate further MFV spurion insertions involving $Y_d$ which
generally lead to mass fine-splittings; for simplicity 
we neglect these terms. The Higgs field
obtains a vacuum expectation value $\langle H \rangle = v/\sqrt{2}$,
breaking the electroweak gauge symmetry and giving a contribution to
the $S$ mass-squared matrix:
\begin{eqnarray}
{\cal L} \supset -  S^*_i \left[ m_A^2 \mathbf{1}_{ij} 
+ m_B^2 (Y_u Y_u^\dag)_{ij}   \right]S_j ,
\label{Smass}
\end{eqnarray}
where we have defined 
\begin{eqnarray}
m_A^2 & = & m_S^2 a + \lambda v^2 a' , \nonumber \\
m_B^2 & = & m_S^2 b+ \lambda v^2 b' .
\end{eqnarray}
Without loss of generality, we freeze the Yukawa spurions to the
background values $Y_d = \lambda_d$, $Y_u=V^\dag \lambda_u$, where
$(\lambda_{u})_{ij}= y_{u,i}\delta_{ij}$, $(\lambda_d)_{ij} =
y_{d,i}\delta_{ij}$ are diagonal in the physical Yukawa couplings $y_{u, i}$
and $y_{d, i}$; quark mass eigenstates are then obtained via $u_L
\rightarrow V^\dag u_L$.
The mass matrix in Eq.~(\ref{Smass}) is diagonalized by $S \rightarrow
V^\dag S$:
\begin{equation}
{\cal L}  \rightarrow   - S^{*}_i \left[ m_A^2 \mathbf{1}_{ij} 
+ m_B^2  (\lambda_u^2)_{ij}    \right]S_j .
\label{spectrum}
\end{equation}
Up to fine splittings induced by up and charm Yukawa couplings, we
find $m_1^2 \simeq m_2^2 \simeq m_A^2$ and $m_3^2 \simeq m_A^2 + m_B^2
y_t^2$.  There are two possibilities for the spectrum: 1) a ``normal''
spectrum with two lighter, nearly degenerate states $S_{1,2}$ and one
heavier state $S_3$, or 2) an ``inverted'' spectrum with two heavier
states $S_{1,2}$ and one lighter state $S_3$.

Of phenomenological importance are the trilinear couplings of scalars
to the Higgs $h$~\cite{singlet}, which in the mass eigenbasis read:
\begin{eqnarray}
{\cal L} & \supset & - 2 \lambda  v h {S}^*_i ( a' \mathbf{1}_{ij} + b' (\lambda_u^2)_{ij} ) S_j \nonumber \\
 & \equiv & - 2 \widetilde \lambda_i  v h {S}^*_i S_i,
\label{trilinear}
\end{eqnarray}
where we have defined 
$\widetilde \lambda_i \equiv\lambda(a'+ b' y^2_{u, i})$
These couplings 
provide annihilation channels into SM particles, and are furthermore
constrained by direct and indirect DM searches. For light DM particles
in the $10-100$ GeV range, the recent XENON100 null
results~\cite{xenon100} imply a limit of $\widetilde\lambda_i \lesssim
0.015~(m_i / 10~{\rm GeV})(m_h / 120\,{\rm GeV})^2$ for the DM
particle.  In addition, the Higgs portal interactions above
lead to nonstandard decay channels of the SM Higgs, $h\rightarrow
S_i^* S_i$, provided that the scalars $S_i$ are light enough. The
partial width for $h\rightarrow S_i^* S_i$ is given by
\begin{equation}
\Gamma_{h\rightarrow S^*_i S_i} = \frac{\widetilde\lambda_i^2 v^2}{4\pi m_h}
\left(1-\frac{4 m_i^2}{m_h^2}\right)^{1/2}.
\label{hSSdecay}
\end{equation}
For values of $\widetilde \lambda_i$ on the order of the $b$-quark
Yukawa coupling $y_b$, these decay modes become competitive with the
SM $h\rightarrow b \bar b$ mode for a light Higgs.

\paragraph{Effective theory.}

By construction, the scalars do not couple to quarks at the
renormalizable level since such operators lead to the decay of the
would-be DM candidate. However, couplings between pairs of DM
multiplets and quarks may occur in an effective theory through higher
dimensional operators. At the dimension six level, we can write the
following effective operators coupling two scalar multiplets to a
quark and anti-quark:
\begin{eqnarray}
{\cal O}^{1}_{ijk\ell} & = & (\bar Q_i \gamma^\mu Q_j )(S_k^*  \overleftrightarrow {\partial_\mu} S_\ell), \label{op1}
 \\
{\cal O}^{2}_{ijk\ell} & = & (\bar u_{R i} \gamma^\mu u_{R j} )(S_k^*  \overleftrightarrow {\partial_\mu} S_\ell), \label{op2}
 \\
{\cal O}^{3}_{ijk\ell} & = & (\bar d_{R i} \gamma^\mu d_{R j} )(S_k^*  \overleftrightarrow {\partial_\mu} S_\ell) ,  \label{op3}
\label{effops}
\\
{\cal O}^{4}_{ijk\ell} & = & (\bar Q_i u_{R j} )(S_k^*  S_\ell) H^\dag  +{\rm h.c.} , \label{op4}
\\
{\cal O}^{5}_{ijk\ell} & = & (\bar Q_i d_{R j} )(S_k^*  S_\ell) H + {\rm h.c.} ,
\label{op5}
\end{eqnarray}
where $i,j,k,l$ are flavor indices. 
From these operators we can construct the effective Lagrangian:
\begin{eqnarray}
{\cal L}_{eff} & = & \frac{1}{\Lambda^2}\sum_{I = 1}^{5} c^{I}_{ijk\ell} {\cal O}^I_{ijk \ell}.
\end{eqnarray}
The coefficients $c^{I}_{ijk\ell}$ in general contain 
all possible flavor contractions consistent with MFV. 
For example, for the operator ${\cal O}^1$ we have 
\begin{eqnarray}
c^1_{ijk\ell}  & = & c^1_1 \mathbf{1}_{ij}  \mathbf{1}_{k\ell} 
+ c^1_2 \mathbf{1}_{i\ell} \mathbf{1}_{kj} 
+ c^1_3 (Y_u Y_u^\dag)_{ij} \mathbf{1}_{k\ell} \nonumber \\ &+& 
c^1_4  \mathbf{1}_{ij}  (Y_u Y_u^\dag)_{k\ell}
+ c^1_5 (Y_u Y_u^\dag)_{i\ell} \mathbf{1}_{kj} \nonumber \\ &+& 
 c^{1*}_5  \mathbf{1}_{i\ell}  (Y_u Y_u^\dag)_{kj} 
+ \dots,
\label{cflavor}
\end{eqnarray}
where $c_i^1$ are Wilson coefficients for a given flavor contraction,
and where we display the leading terms in the MFV expansion up to one
insertion of $Y_u Y_u^\dag$.

Such higher-dimension operators can have a significant impact on the
cosmology by providing the DM with efficient annihilation channels to
SM quarks, as long as the effective scale $\Lambda/\sqrt{ {c^{I}_i} }$
suppressing the operators is not too much larger than the weak scale.
The same operators can potentially be probed by direct and indirect
detection experiments, colliders, and precision flavor studies.

Indeed, certain operators and flavor structures are more constrained
than others. To illustrate this fact, consider for concreteness the
``inverted'' spectrum from Eq.~(\ref{spectrum}) in which the DM
particle is $S_3$. For the operator ${\cal O}^1$, the flavor
structures associated with the coefficients $c_1^1$ and $c_4^1$ in
Eq.~(\ref{cflavor}) lead to the vector coupling of the DM particle
$S_3$ to valence quarks in the nucleon, and are therefore strongly
constrained by direct detection experiments. On the other hand, the
flavor structures associated with $c_2^1$, $c_3^1$, and $c_5^1$ imply
that the vector couplings of the DM state $S_3$ are dominantly to the
third generation quarks, while the couplings to valence quarks either
vanish or are Yukawa and/or CKM suppressed. Therefore, these
coefficients are far less constrained by direct DM searches.

Continuing on with this example, there is the possibility of new FCNCs
induced by the operator ${\cal O}^1$ once the DM multiplet is
integrated out. Whereas the coefficients $c_1^1$ and $c_4^1$ in
Eq.~(\ref{cflavor}) lead to strictly flavor-diagonal couplings in the
quarks, the flavor structures associated with $c_2^1$, $c_3^1$, and
$c_5^1$ indeed do induce FCNCs in the down-quark sector due to the
presence of CKM mixing.

A number of other phenomenological considerations must be taken into
account, each of which potentially constrain the Wilson coeffcients
$c^I_{ijkl}$ for the operators in
Eqs.~(\ref{op2}-\ref{op5}). Depending on the UV completion some of
these coefficients may be sizable while others may be suppressed or
vanish. From the effective theory perspective, one can explore the
consequences of each operator on the cosmology and phenomenology in a
model-independent way. Understanding which operators are constrained
and which are allowed may then point towards
the underlying dynamics at scales greater than~$\Lambda$.
A comprehensive analysis of the effective operators
(\ref{op1}-\ref{op5}) and their various flavor structures will be left
for future work. Instead, we shall specialize to one particular
operator and examine in detail the associated cosmology and
phenomenological constraints and prospects.

For the remainder of the paper let us consider operator ${\cal O}^5$
in Eq.~(\ref{op5}) with the following flavor structure:
\begin{equation} {\cal L}_{\rm eff} = \frac{c}{\Lambda^2}[ \bar Q_i S_i ][
  S^*_j (Y_d)_{jk} d_{R k} ] H+ {\rm h.c},
\end{equation}
where the flavor indices are contracted within the brackets, while the
Lorentz, color, and $SU(2)_L$, indices are contracted outside.
The coefficient $c$ is taken here to be a phase, so that $|c| =1$.
As we will demonstrate, this operator allows for a viable cosmology while being consistent with a variety of constraints.

After electroweak symmetry breaking and diagonalization of the quark
and scalar masses, we are left with the following effective
Lagrangian:
\begin{equation} {\cal L}_{\rm eff} \rightarrow
  \frac{c}{\Lambda^2}\frac{v}{\sqrt{2}}[ \bar d_{L i} V^\dag_{ij} S_j
  ][S^*_k (V \lambda_d)_{k\ell} d_{R \ell} ] + {\rm h.c.}~.
\label{operator}
\end{equation}
Focusing on the ``inverted'' spectrum below, $m_3 < m_1\simeq m_2$,
$S_3$ is the stable DM candidate, while $S_1$ and $S_2$ are unstable.

\begin{figure}\centerline{
\includegraphics[width=\columnwidth]{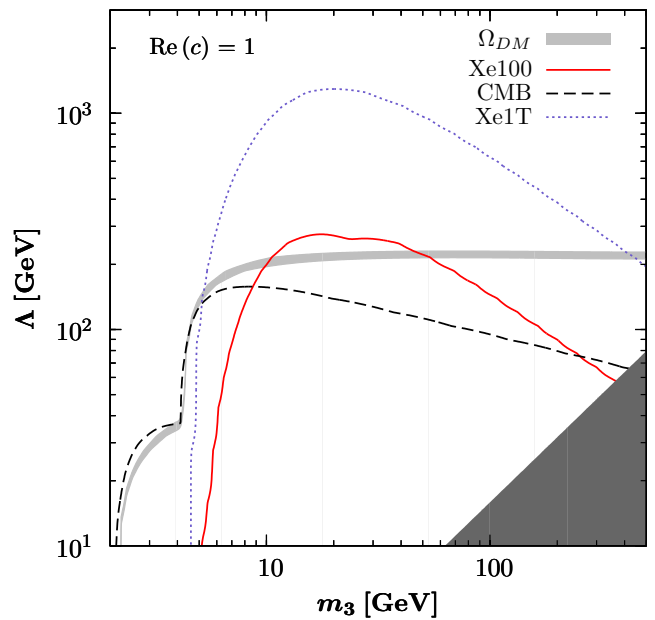}
}
\caption{\small {Constraints:} Shown are \textit{lower} limits on the
  effective scale $\Lambda$ as a function of $m_3$ for
  $\mathrm{Re}(c)=1$. The dashed line is a constraint on the
  annihilation cross section from CMB temperature and polarization
  anisotropies. The solid (red) line is the exclusion bound from the
  recent XENON100 direct detection experiment with 100 live-days of
  exposure; the dotted (blue) line shows the sensitivity of a future
  ton-scale liquid xenon experiment. In the lower right shaded region,
  the effective field theory description breaks down. Within the gray
  shaded band, the relic density of $S_3$ matches the observed DM
  density, $\Omega_{S_3+S_3^{*}} h^2 = \Omega_{DM} h^2$. }
\label{fig:DD}
\end{figure}

\paragraph{Relic abundance.} 
First we consider the thermal relic abundance of $S_3$. If the DM mass
$m_3$ is larger than the bottom quark mass $m_b$, the dominant
annihilation mode is $S_3 S_3^* \rightarrow \bar b b$, coming from the
following term contained in Eq.~(\ref{operator}):
\begin{eqnarray}
  \mathcal{L}_{\rm eff} & \supset & \frac{c}{\Lambda^2} m_b |V_{tb}|^2 S_3^*S_3 \bar b_L b_R +{\rm h.c.}\, ,
\label{33bb}
\end{eqnarray}
where $|V_{tb}| \simeq 1 $ is the top-bottom CKM matrix element.
The above term in the Lagrangian leads to the following thermally
averaged annihilation cross section:
\begin{eqnarray}
\langle \sigma v \rangle_{33 \rightarrow \bar b b} & = & 
 \frac{3 }{4\pi \Lambda^4} m_b^2 |V_{tb}|^4 \left( 1-\frac{m_b^2}{m_3^2} \right)^{1/2}  \\
&\times &\left\{ [{\rm Re}(c)]^2 \left(1-\frac{m_b^2}{m_3^2}\right) +[{\rm Im}(c)]^2  \right\}.
\nonumber 
\end{eqnarray}
In the limit $m_3 \gg m_b$, the annihilation cross section
approximately scales as
\begin{equation}
\langle \sigma v \rangle_{33 \rightarrow \bar b b} \simeq1\,{\rm pb} \left( \frac{200\,\GeV}{\Lambda}\right)^4,
\end{equation}
suggesting that an effective scale of $\Lambda \approx 200$ GeV is
required in order to match with the nominal $3\sigma$ DM abundance
$\Omega^{3\sigma}_{DM}h^2 = 0.1109 \pm 0.017 $ as inferred from the
seven year data sample of the WMAP satellite
experiment~\cite{Komatsu:2010fb}. This can also be seen in
Fig.~\ref{fig:DD} where the gray shaded band indicates the region in
which $S_3+S_3^{*}$ attain the right relic abundance. Below the
threshold of $b\bar b$ annihilation, $m_3 \lesssim 4.2\,\GeV$, the
most efficient channel is $S_3^{*}S_3 \rightarrow s\bar b, b \bar s$
with $\langle \sigma v \rangle \propto |V_{ts}|^2 |V_{tb}|^2 m_b
m_s/\Lambda^4 $.

\paragraph{CMB constraints.}
Residual annihilation of $S_3$-pairs has its strongest impact at
cosmic redshifts $z\sim 1000$ at the time of recombination of
electrons with hydrogen (and helium.) The electromagnetic cascades
formed in the annihilation delay recombination thus affecting the
temperature and polarization anisotropies of the cosmic microwave
background radiation (CMB)~\cite{Slatyer:2009yq,Hutsi:2011vx}.  Since
the rate of injected energy scales as $ \langle \sigma v \rangle
n_{DM}^2 m_{DM} \propto \langle \sigma v \rangle / m_{DM}$ where
$n_{DM}$ is the DM number density, the CMB limit on the annihilation
cross section can conveniently, and in a rather model-independent
fashion, be expressed in the form
\begin{equation}
  \label{eq:cmb}
  (1-f_{\nu}) \langle \sigma v \rangle / m_{3} < r . 
\end{equation}
Here, $f_{\nu}$ is the fraction of energy carried away by neutrinos
and we employ $r = 0.191\,{\rm GeV}/{\rm cm^3 s^{-1}}$ as obtained
in~\cite{Hutsi:2011vx}. Using the PYTHIA~\cite{Sjostrand:2006za} model
for $e^{+}e^{-}$ annihilations to hadrons we have verified that
$f_{\nu}$ always remains below $10\%$; as a representative value we
choose $f_{\nu} = 0.08$. The resulting lower limit on the effective
scale $\Lambda$ is shown by the dashed line in Fig.~\ref{fig:DD} and
is strongest for small values of $m_3$, excluding the otherwise
cosmologically viable region $m_3\lesssim 5\,{\rm GeV}$.

\paragraph{Direct detection.}

The effective operator (\ref{operator}) also induces the elastic
scattering of $S_3$ on nuclei. The following terms lead to
spin-independent scattering:
\begin{equation}
  \mathcal{L}_{\rm eff}   \supset 
  \frac{{\rm Re}(c)}{\Lambda^2}
  \sum_{i=1}^3 m_{d_i} |V_{3i}|^2 S_3^* S_3 \bar d_i d_i . 
\label{33qq}
\end{equation}
Though the couplings have the usual dependence on the quark masses, as
for example expected in Higgs-mediated DM scattering on nuclei, the
couplings to light quarks $s$ and $d$ are additionally CKM
suppressed. This implies that only the scattering of $S_3$ on the
$b$-quark content of the nucleons is relevant in constraining the
scale $\Lambda$ as well as the phase $c$ of the higher-dimensional
operator~(\ref{operator}).
The nucleon  matrix element under question reads
\begin{equation}
  f_{n,b} \equiv \langle n| m_b \bar b b | n\rangle = m_n \frac{2}{27} f^{(n)}_{TG}  \simeq 0.04,
\end{equation}
where we have used that the fact that the $b$-quark matrix element can
be expressed in terms of the matrix element of the gluon scalar
density~\cite{shifmanDM}, 
$f^{(n)}_{TG} = 1 - \sum_{u,d,s} f^{(n)}_{Tq} \simeq
0.57$~\cite{Ellis:2008hf}.

This leads to the effective spin-independent elastic DM-nucleon cross
section,
\begin{eqnarray}
  \sigma_n & = & \frac{  [{\rm Re}(c)]^2 |V_{tb}|^2  f^{2}_{n,b}\, \mu_n^2 }{4 \pi m_3^2 \Lambda^4}  \\
  & \simeq &
  3 \times 10^{-43}\,{\rm cm}^2\,  [{\rm Re}(c)]^2  \left(\frac{10\,\rm GeV}{ m_3}\right)^2 \left( \frac{200\,\rm GeV}{\Lambda} \right)^4 . \nonumber 
\label{sign}
\end{eqnarray}

The upper limit on $\sigma_n$ from direct detection experiment can be
translated into a lower limit on $\Lambda$. The most constraining data
in this respect comes from the recent 100~live-day result of the
XENON100 collaboration~\cite{xenon100}.  We obtain the constraint by
using Yellin's maximum gap method~\cite{Yellin:2002xd} which accounts
for the 3 observed events within the 48~kg fiducial detector volume
and employ the results of~\cite{Sorensen:2010hq} to account for
detector resolution, efficiency, and acceptance. The resulting limit
in the $(m_{DM},\sigma_n)$ plane is in very good agreement with the
one of~\cite{xenon100}. For the sake of exploring future sensitivity
of direct detection experiments we also simulate a ton-scale liquid
xenon detector assuming a raw exposure of 1~ton$\times$year using the
detector model for XENON10 of Ref.~\cite{Sorensen:2010hq}.

Choosing $\mathrm{Re}(c)=1$, the XENON100 lower limit on $\Lambda$ as
a function of $m_3$ is shown by the solid (red) line in
Fig.~\ref{fig:DD}. As can be seen, mass values in the range
$10\,\mathrm{GeV}\lesssim m_3 \lesssim 60\,\mathrm{GeV}$ are
excluded. The dotted (blue) line illustrates that a future ton-scale
direct detection experiment may well be sensitive to the entire
electroweak-scale parameter region of $m_3$, excluding values up to
$m_3\lesssim 400\,\mathrm{GeV}$. It is important to note, however,
that since the overall phase $c$ of the higher dimensional operator is
a free parameter, one can avoid completely the direct detection
constraints by choosing $\mathrm{Re}(c)\ll1$ while still obtaining the
correct relic abundance. Finally, the the shaded region in the lower
right corner of Fig.~\ref{fig:DD}---subject to the condition $\Lambda
< m_3/2\pi$---indicates the part of the parameter space that does not admit a
perturbative UV-completion~\cite{mono}.

\paragraph{Flavor physics.}
\begin{table*}[t]
\begin{center}
\renewcommand{\arraystretch}{1.3}
 \begin{tabular}{ | c | c | c | c | c | c | c | c  | c| c| c| c| c|}
\hline
  $m_h$ &$m_{1}$  &  $m_{2}$  &  $m_{3}$ 
 &  $\widetilde \lambda_{1}$ &   $\widetilde \lambda_{2}$      &   $\widetilde \lambda_{3}$     
 &   $m_S^2 $      &   $\lambda $      &   $a$      & $a'$ & $b $  & $b'$
  \\
 \hline \hline
120 &  25  & 25  & 10  & 0.15 & 0.15 &  0.01   & $-86^2$ & $0.35$ &  1.14   &  0.429  & -1.10  & -0.409
\\  \hline     
    \hline
  \end{tabular}
\end{center}
\caption{  Benchmark spectrum and couplings for LHC study. The input parameters are $m_{1}$, $m_2$, $m_3$, $\widetilde \lambda_1$,  $\widetilde \lambda_2$,  $\widetilde \lambda_3$, $m_S^2$ and $\lambda$. This determines the Lagrangian parameters  $a$, $a'$, $b$, $b'$. All mass parameters are in GeV.   }
\label{benchmark}
\end{table*}

The new scalars can lead to new FCNCs at the one loop level. Here we
will focus on the contribution to $K^0$--$\bar K^0$ mixing.  The relevant
interactions from the operator (\ref{operator}) that can in principle
induce $K^0$--$\bar K^0$ mixing are
\begin{eqnarray}
{\cal L}_{\rm eff} & \supset &\sum_{i,j} \frac{c}{\Lambda^2} \,m_s V^\dag_{1i} V_{j2} S^*_j S_i \bar d_L s_R  + {\rm h.c.}~ .
\label{KKterms}
\end{eqnarray}
Integrating out the scalars one arrives at the following Lagrangian:
\begin{equation}
  \mathcal{L}_{sd} = C^{sd}_S (\bar s_R d_L)(\bar s_R d_L) + {\rm h.c.}~,
\label{KKeff}
\end{equation}
where the Wilson coefficient is given by
\begin{equation}
C^{sd}_S \simeq \frac{(c^*)^2 m_s^2}{32 \pi^2 \Lambda^4} 
(V_{td} V^*_{ts})^2  F\left(\frac{m_3^2}{m_1^2} \right) ,
\label{wilson2}
\end{equation}
with $F(\hat x) = (\hat x +1)(\hat x-1)^{-1}\log{\hat x}-2$.  The
function $F(\hat x)$ takes ${\cal O}(1)$ values for typical values of
the masses, and vanishes for $\hat x = 1$. To arrive at
Eq.~(\ref{wilson2}) we have made use of the unitarity of the CKM
matrix and assumed a mass splitting predicted by MFV~(\ref{spectrum}). With $\Lambda =
{\cal O}(v)$ as dictated by the relic abundance calculation, we
observe that the operator $(\ref{operator})$ leads to a contribution
in Eq.~(\ref{wilson2}) which is suppressed compared to the SM value by
a factor $m_s^2/m_W^2 \approx 10^{-6}$. Therefore, we are comfortably
below the experimental limits.  Similarly, for $B_q - \bar B_q$
mixing, we find operators that are suppressed compared to the SM value
by a factor $m_b^2/m_W^2 \approx 10^{-3}$. Furthermore, we find that
no contribution to the inclusive decay $b \rightarrow s \gamma$ is
induced.

\paragraph{Decays of heavy dark flavors.}

The heavier scalars $S_1$ and $S_2$ are unstable and decay via the
higher dimensional operator~(\ref{operator}). The leading operators
which are unsuppressed in powers of light quark masses $m_{s,d}$ and
off-diagonal CKM elements are
\begin{eqnarray} {\cal L}_{\rm eff} & \supset & \frac{c}{\Lambda^2}
  m_b V_{tb} V^*_{cs} S_3^* S_2 \bar s_L b_R \nonumber \\ & +&
  \frac{c}{\Lambda^2} m_b V_{tb} V^*_{ud} S_3^* S_1 \bar d_L b_R +
  {\rm h.c.}~.
\end{eqnarray}
These operators mediate the decays
\begin{eqnarray}
\label{heavydecay1}    S_2 & \rightarrow &S_3 s \bar b ,\\ 
\label{heavydecay2}     S_1 & \rightarrow & S_3 d \bar b.
\end{eqnarray}
In the limit 
$m_{1,2} \gg m_b,m_3$, the three-body partial decay
width for $S_i\rightarrow S_3 q \bar b$ is approximately given by
\begin{eqnarray}
\Gamma_{i\rightarrow 3 q b} & \simeq & 
\frac{  |c V_{tb} V^*_{ii}|^2 m_b^2 m_i^3 }{ 512 \pi^3 \Lambda^4} \\ 
& \simeq & 10^{-5}\, {\rm MeV} \times \left(  \frac{m_i}{25\, \rm GeV}\right)^{3}
\left(  \frac{200\, \rm GeV}{\Lambda}\right)^{4} .\nonumber
\end{eqnarray}
The scalars therefore decay promptly.

\paragraph{LHC prospects.}

A novel prediction of our flavored DM scenario is the presence of  heavy dark flavors. 
Given a production mechanism, such heavy flavors can lead to interesting
signatures at the Large Hadron Collider (LHC).  The effective operator (\ref{operator}) can in principle lead to pair production via the parton-level processes $q_i \bar q_j \rightarrow S_i S_j^*$. In practice, however, the production cross section for these processes is negligible due to the Yukawa suppressed couplings in Eq.~(\ref{operator})  (this is also why the operator faces no bounds from monojet searches at the Tevatron~\cite{mono}). 

Here we instead focus on
production of $S_{1,2}$ via the Higgs portal interaction in
Eq.~(\ref{portal}). If the new scalar states are lighter than $m_h/2$ and
have a trilinear coupling to the Higgs in Eq.~(\ref{trilinear}) larger
than the bottom Yukawa interaction $y_b$, then
the Higgs decays predominantly to intermediately heavy scalars
$S_{1,2}$ with a partial width given by Eq.~(\ref{hSSdecay}). These
scalars subsequently decay via the three-body processes
(\ref{heavydecay1},\ref{heavydecay2}) to a bottom quark, a light quark
($d$ or $s$), and the DM candidate $S_3$,
\begin{eqnarray}
\label{LHCmode1}
 h & \rightarrow & S^*_{1} S_{1} \rightarrow b\bar{b} d\bar{d} S^*_3 S_3, \\
\label{LHCmode2}
 h & \rightarrow & S^*_{2} S_{2} \rightarrow b\bar{b} s\bar{s} S^*_3 S_3.
\end{eqnarray}
We now investigate the potential of the LHC ($\sqrt{s}=14$ TeV) to discover
the processes (\ref{LHCmode1},\ref{LHCmode2}) for the benchmark scenario given in
Table~\ref{benchmark}.  For this choice of parameters we find that the
Higgs decays almost exclusively via (\ref{LHCmode1},\ref{LHCmode2}). 
  The most promising channel to disentangle the signal
from the large SM backgrounds is 
expectedly the $HZ$ production channel, where the $Z$ boson
subsequently decays to electrons or muons. One immediate reason for
this is that the resulting hard leptons provide a triggerable signal.
Furthermore, if the $Z$ boson is boosted and the Higgs recoils against
the leptonically decaying $Z$, the leptons yield a good handle to
calibrate the 
jet energy and to measure the amount of $\not \!\! E_T$ in the
event. Similar configurations have been used in previous analyses for
a boosted SM Higgs boson which decays into a bottom quark
pair~\cite{hzboosted}.

We simulate the signal using MadEvent \cite{Alwall:2007st} and the
backgrounds using Herwig++ \cite{Bahr:2008pv}. The NLO production
cross sections for the backgrounds and the signal was calculated using
MCFM \cite{mcfm}.
To select a lepton we demand it to be central and sufficiently hard,
\begin{alignat}{5}
|y_l| < 2.5, \qquad \qquad \qquad 
p_{T, l} &> 20~\text{GeV} . 
\label{eq:rap}
\end{alignat}
For the hardest lepton in the event we further require
\begin{alignat}{5}
p_{T, l_1} &> 80~\text{GeV} . 
\end{alignat}
The muons and electrons must be isolated, \textit{i.e.}, the ha\-dronic
transverse energy in a cone of $R=0.3$ around the lepton has to be
$E_{T_{\text{hadronic}}} < 0.1~E_{T, l}$. We accept
events with exactly two isolated leptons.
The $Z$ bosons are reconstructed by combining two oppositely charged
isolated electrons or muons, requiring
\begin{equation}
\label{eq:mz}
m_Z-10~\text{GeV} < m_{ll} < m_Z+10~\text{GeV}
\end{equation}
and $p_{T,ll}> 150~\text{GeV}.$

\begin{figure}\centerline{
\includegraphics[width=\columnwidth]{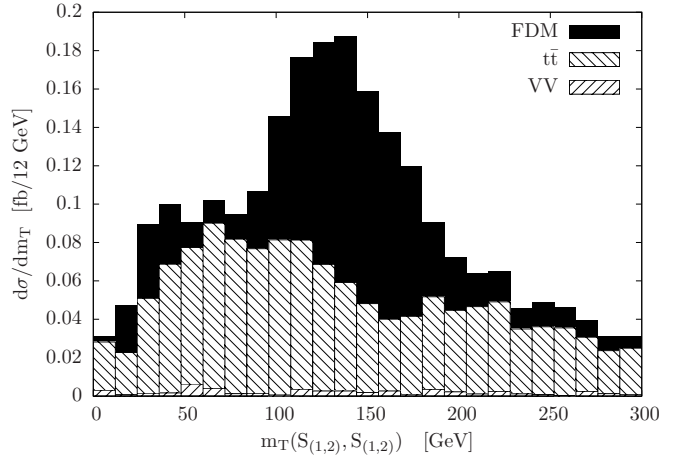}
}
\caption{\small Reconstruction of $m_T$ according to Eq.~(\ref{eq:mT}) for a Higgs mass of 120 GeV. In addition to the cuts outlined in Eqs.~(\ref{eq:rap})-(\ref{eq:dphi}) we require at least two jets to accept an event.}
\label{fig:mT}
\end{figure}

We group the remaining visible final state particles into ``cells'' of
size $\Delta \eta \times \Delta \phi = 0.1 \times 0.1$. All particles
in a cell are combined, and the total three-momentum 
is rescaled such that each cell has zero invariant mass. Cells with
energy $< 0.5~\rm{GeV}$ are discarded, while the remaining ones are
clustered into jets. We recombine the jets using Fastjet
\cite{fastjet} with the anti-kT jet algorithm \cite{antikt}, $R=0.7$,
and
\begin{equation}
  p_{T,j} > 30~\rm{GeV}.
\end{equation}
We reject events where a jet with $p_{T,j}>50~\rm{GeV}$ has a smaller angular separation to the Z boson than $\Delta R_{j,Z}<1.5$.
At least one jet must be present in the event, and the hardest jet in
the event must be $b$-tagged. For the $b$-tag we assume an efficiency
of $60\%$ and a fake rate of $2\%$.  The $S_3$ states escape the
detector unobserved. Therefore we accept an event if
\begin{equation}
  \not \!\! E_T   > 50~\rm{GeV},
\end{equation}
and calculate $\not\!\!  E_T$ from all visible objects with $|y| < 5.0$\,.
The resulting $\not\!\!\! E_T$ vector must have a large azimuthal separation from
the reconstructed $Z$ boson,
\begin{equation}
\label{eq:dphi}
\Delta \phi_{~\!{\not  E_T,Z}} > 2.0.
\end{equation}
With these cuts we obtain $S/B \simeq 1$ and $S/\sqrt{B} \simeq 5$
after $15 ~\rm{fb}^{-1}$; see Table \ref{crosssecs} for
details. Therefore an excess in events with $\not \!\! E_T +
b\bar{b}+l^+l^-$ can be observed relatively early at the LHC.
However, due to the
sizable fraction of missing transverse energy, $\not \!\!E_T$, the
reconstruction of the mass of the Higgs boson is
challenging.
Because the bottom quark and the light quark from the three-body-decay
of $S_{1,2}$ are highly collimated they will be observed as one
jet. Thus, after requiring at least two jets in the event, in the rest
frame of the Higgs one finds $m_{S_3^*S_3} \simeq m_{jj}$. Following
\cite{dieter}, we use
\begin{equation}
\label{eq:mT}
m_T(S^*_{i}S_{i})=\sqrt{(\not \!\! E'_T+E_{T,jj})^2-({\bf p}_{T,jj}+{\bf \not \!\! p}_T)^2 },
\end{equation}
to reconstruct the Higgs mass as shown in Fig.~\ref{fig:mT}. In
Eq.~(\ref{eq:mT}), the transverse energies are given by ${\not \!\!
  E}'_T=\sqrt{{\bf \not \!\! p}_T^2+m^2_{jj}}$ and
$E_{T,jj}=\sqrt{{\bf p}^2_{T,jj}+m^2_{jj}}$. $m_{jj}$ is the invariant mass of the two hardest jets in the event which have a large angular separation to the $Z$ boson, $R_{j,Z}>2.0$. 
For a rough mass
reconstruction of the Higgs boson in this channel we expect
$\mathcal{O}(100)~\rm{fb}^{-1}$, but other Higgs production channels
could be added.  If $m_{1,2} \lesssim m_3 + m_b$ the Higgs decays to
jets and missing transverse energy. This signature displays strong
similarities with the so-called buried Higgs models
\cite{burried_higgs}. In either case, jet substructure can help to
reconstruct the Higgs boson \cite{sub_higgs}.

%
\begin{table*}[t]
\begin{tabular}{c || c | c | c | c | c }
  & FDM  & $t \bar{t}$ & $ZZ$ & $WZ$ & $WW $ \\	
\hline
$n_j \geq 1$, $n_l=2$ and $p_{T,l_1}>80~\rm{GeV}$    & 12.7	& 8903.7 	& 202.3 	& 168.5  & 242.2 \\
$\not \!\! E_T   > 50~\rm{GeV}$ 			& 7.8 	& 5744.1 	& 20.6 	& 20.4    & 118.8   \\
$Z$ reconst. and $p_{T,Z}>150~\rm{GeV}$, no $\Delta R_{j_{50},Z} < 1.5 $	& 4.3 	& 9.9 	& 5.8 	& 3.8      & 0.7  \\
$ \Delta \phi_{~\!{\not \! E_T,Z}} > 2.0 $	& 4.2 	& 4.6 	& 5.2 	& 3.3      & 0.03 \\
b-tag & 2.2 	& 2.2 	& 0.2 	& 0.1 	& 0.01  \\
\end{tabular}
\caption{Cross sections for different steps in the analysis for the signal 
  from the flavored dark matter model (FDM) and the dominant SM backgrounds. 
  Systematic uncertainties in the measurement of $ \not \!\! E_T$ can elevate 
  $Z$+jets to a relevant background but their simulation is beyond the scope of this work. All cross sections are normalized to NLO QCD precision, calculated using MCFM, and are given in femtobarn. The background events were generated using Herwig++ and the signal was generated using MadEvent and showered using Herwig++.}
\label{crosssecs}
\end{table*}

\paragraph{\bf Discussion.}

In this paper we have demonstrated that the MFV hypothesis, motivated
by the need to suppress dangerous flavor changing processes in new
physics scenarios, also provides a novel organizing principle for the
physics of DM.  Indeed, MFV leads to the stability of the lightest
neutral particle for a large number of flavor representations. We have
determined which representations have stable particles.
This opens up the possibility for a large class of new flavored DM
candidates.

Because of the connection with flavor physics, the phenomenology of
these models is very rich.  In addition to providing stability, MFV is
a predictive framework for DM. The mass spectrum of the flavored DM
multiplet is governed by the hierarchical nature of the quark Yukawa
couplings. Furthermore, the pattern of couplings between the DM and
the SM is prescribed by the flavor symmetries of $G_q$ and CKM
breaking. Our example study of the effective theory of a flavor
$SU(3)_Q$ triplet, gauge singlet scalar DM illustrates many of the
general aspects of the cosmology and phenomenology, and leads to a
viable DM scenario with testable predictions at future direct
detection experiments and present day colliders. In particular, the
signatures of the heavy unstable dark flavors are quite novel and can
be searched for at the LHC. However, it is fair to say that we have
only scratched the surface of the phenomenological implications of the
flavored DM scenario.

In this paper we have focused on the flavor symmetries of the quark sector, primarily 
because of our still incomplete knowledge about the SM neutrino sector. However, with some assumptions about how neutrinos acquire mass the MFV hypothesis can be formulated for leptons~\cite{LMFV}, and it would therefore be interesting to investigate the possibility of leptonically flavored DM. In this case, it is conceivable that heavy dark flavors may leave their imprint via decays to DM with associated leptons.

It would be interesting investigate more generally the predictions for
other flavored DM multiplets, including larger flavor representations
and electroweak multiplets.  For the latter, typically because of the
efficient annihilation to SM gauge bosons, the DM particle must be
quite heavy to be in accordance with the observed cosmic
abundance~\cite{minimalDM}.  Multiple flavors of stable electroweak DM
may instead be comparatively light since each flavor then only
constitutes a portion of the DM density.

More comprehensive studies of flavored DM effective theories is
certainly warranted. Much attention as of late has been given to
effective field theories of DM, notably as a means to investigate the
potential of hadron colliders to probe DM-quark interactions in a
manner complementary to direct detection experiments~\cite{mono}. In
these studies, the couplings of DM to quarks were taken to be flavor
universal for vector couplings and proportional to mass for scalar couplings to avoid FCNC constraints. Of course, a less restrictive
possibility is to abide by the MFV principle, which allows for more
general couplings between DM and quarks, including flavor violating
couplings governed by the CKM elements.

Ultimately, since flavored multiplets cannot couple to quarks at the
renormalizable level (else they would decay), it is interesting to
construct UV completions to such effective theories of flavored
DM. One may postulate additional particles that connect the DM
multiplet to SM quarks, which themselves must also obey MFV. Indeed it
is certainly interesting to speculate about a new sector of MFV
matter, given that MFV has proven useful in addressing recent Tevatron
anomalies, such as the like-sign dimuon asymmetry in
$B$-decays~\cite{dimuonEX,dimuonTH} and the top quark forward-backward
asymmetry~\cite{topEX,topTH}. Perhaps, the DM particle is contained in
such an MFV sector.

The evidence for DM is one of many reasons to expect new physics, and
yet the SM remains impeccable when confronted with experiment. This
motivates new symmetry principles designed to forbid dangerous
interactions of new particles, which may also have the dual purpose of
stabilizing DM.
The canonical example of such a symmetry is a discrete $Z_2$ parity.
In this paper, we have shown that flavor symmetries together with the
MFV principle provide another compelling possibility, thus pointing the
way towards new avenues for physics beyond the SM 
and DM.
\vspace{5pt}

\subsubsection*{\bf Acknowledgements}

We thank Uli Nierste, Nilendra Deshpande, and Maxim Pospelov for helpful
discussions. BB is grateful to Jure Zupan for stimulating discussions about his related work~\cite{juretalk}. Research at the Perimeter Institute is supported in part
by the Government of Canada through NSERC and by the Province of
Ontario through MEDT. MS was supported by US Department of Energy
under contract number DE-FG02-96ER40969. 
MS thanks the Perimeter Institute for hospitality.

\end{document}